\documentclass[prl,twocolumn,showpacs,superscriptaddress,amsmath,amssymb,reprint]{revtex4}

\usepackage{graphicx}
\usepackage{latexsym}
\usepackage{amsmath}
\usepackage{amssymb}
\usepackage{amsfonts}
\usepackage{color}

\newcommand{\ds}{\displaystyle}

\begin{document}
\bibliographystyle{apsrev}

\title{Double path interference and magnetic oscillations in Cooper pair transport through a single nanowire}

\author{S. V. Mironov}
\affiliation{University Bordeaux, LOMA UMR-CNRS 5798, F-33405 Talence Cedex, France}
\affiliation{Institute for Physics of Microstructures, Russian Academy of Sciences, 603950 Nizhny Novgorod, GSP-105, Russia}
\author{A. S. Mel'nikov}
\affiliation{Institute for Physics of Microstructures, Russian Academy of Sciences, 603950 Nizhny Novgorod, GSP-105, Russia}
\affiliation{Lobachevsky State University of Nizhny Novgorod, 23 Gagarina, 603950, Nizhny Novgorod, Russia}
\author{A. I. Buzdin}
\affiliation{University Bordeaux, LOMA UMR-CNRS 5798, F-33405 Talence Cedex, France}

\date{\today}
\begin{abstract}
We show that the critical current of the Josephson junction consisting of superconducting electrodes coupled through a nanowire with two conductive channels can reveal the multi-periodic magnetic oscillations. The multi-periodicity originates from the quantum mechanical interference between the channels affected by both the strong spin-orbit coupling and Zeeman interaction. This minimal two-channel model is shown to explain the complicated interference phenomena observed recently in Josephson transport through Bi nanowires.
\end{abstract}


\maketitle

The systems with a few conductive channels are in the focus of the current research in the field of nanoelectronics. Such structures provide unique possibility to construct the devices with tunable transport properties at the quantum length scale. One of the promising realizations of these devices is based on the localized electronic states appearing, for example, at the surface of topological insulators \cite{TI}, at the edges of graphene nanoribbons \cite{Nanoribbons}, and different types of nanowires \cite{EPG,Mourik,Nikolaeva}. The physics of the charge transport through these states appears to be extremely rich due to the strong spin-orbit coupling, large anisotropic g-factors, etc. The wide range of possible non-trivial phenomena stimulates both theoreticians and experimentalists in their research of the edge states and the search of new perspective applications. In particular, the growing interest is attracted to the physics of the edge states coupled to the bulk superconducting leads \cite{TI,EPG}. Such coupling provides a possibility to construct new type of Josephson devices with controllable current-phase relations \cite{Buzdin_RMP,Buzdin_Phi} and favorable conditions for observation of Majorana fermions \cite{Alicea}. An obvious way to get an insight into the properties of these systems is to apply an external magnetic field $H$ and study the resulting dependence of the Josephson critical current vs $H$. Moreover, the magnetic field in such a setup can be used as a tool of effective control of the current-phase relation.

It is the goal of this Letter to describe the magnetotransport phenomena in a Josephson system containing a few conductive channels modeling the edge states localized, e.g., at the surface of a single nanowire. Our consideration is based on the generic model accounting for only two interfering electron paths or conductive channels and strong spin-orbit and Zeeman interactions mentioned above. This model allows to describe both orbital and spin mechanisms of the magnetic field effect as well as non-trivial ground state of the Josephson junction with non-zero superconducting phase difference. The Zeeman interaction produces the spatial oscillation of the Cooper pair wave-function at the scale $\hbar v_F/g \mu_B H$ (similar to the ones in superconductor-ferromagnet structures \cite{Buzdin_RMP}) which result in the magnetic oscillations in the critical current with the characteristic period $\hbar v_F/g \mu_B L$, where $L$ is the channel length. The orbital effect causes a standard phase-gain $\sim 2\pi HS/\Phi_0$ ($\Phi_0=\pi\hbar c/\left|e\right|$ is the flux quantum) in the electronic wave-function similar to the one appearing in the Aharonov-Bohm (AB) effect. Here $S$ is the area enclosed by the pair of the interfering paths projected on the plane perpendicular to the magnetic field. The interfering quantum mechanical amplitudes in this case cause the magnetic oscillations in the total transmission amplitude with the period $2\Phi_0/S$. The Andreev reflection at the superconducting boundaries can double the effective charge in the oscillation period \cite{Cayssol} and, thus, the resulting critical current in the general case oscillates with the competing periods $2\Phi_0/S$ and $\Phi_0/S$. This physical picture should be, of course, modified in the presence of the spin-orbit coupling which is responsible for the dependence of the Fermi-velocity on spin projection and momentum direction. Such specific dependence produces the spontaneous Josephson phase difference \cite{Buzdin_Phi,Reynoso,Krive} and can cause substantial renormalization of the above oscillation periods.

Turning to the existing experimental data we must note that the multi-periodic magnetic oscillations have been recently observed in measurements of the Josephson critical current through the Bi nanowires \cite{Bouchiat_ArXiv}. Such wires are known to reveal the unusual combination of properties mentioned above: (i) strong Rashba spin-orbit coupling with the energy comparable with the Fermi energy \cite{Koroteev,Hirahara}; (ii) large $g$-factor $\sim 10^2$ for certain directions of magnetic field \cite{Seradjeh}; (iii) large Fermi wavelength $\lambda_F\sim 50~{\rm nm}$ \cite{Hofmann}, which makes it easy to create nearly one-dimensional wires. As we show below our model can provide a simple fit of the oscillatory behavior discovered in \cite{Bouchiat_ArXiv} being, thus, a promising candidate for the description of the interference physics in such systems.

%%%%%%%%%%%%%%%%%%%%%%%%%%%%%%%%%%%%%%%%%%%%%
\begin{figure}[t!]
\includegraphics[width=0.4\textwidth]{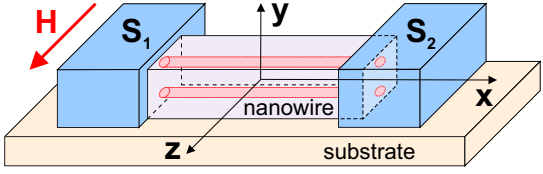}
\caption{(Color online) A model Josephson junction with a two-channel nanowire in external magnetic field.} \label{Fig_System}
\end{figure}
%%%%%%%%%%%%%%%%%%%%%%%%%%%%%%%%%%%%%

We now proceed with the calculation of the critical current of the two-channel nanowire within the Bogoliubov-de Gennes (BdG) approach.
We assume the nanowire of the length $L$ to be placed on top of the insulating substrate and put in contact with two superconducting leads $S_1$ and $S_2$ with the gap functions $\Delta_s e^{-i\varphi/2}$ and $\Delta_s e^{i\varphi/2}$, respectively (see Fig.~\ref{Fig_System}). We choose the origin of the Cartesian coordinate system at the middle of the wire. The $x$-axis is taken along the wire and the $y$-axis is chosen perpendicular to the substrate surface. The parallel conductive channels pass along the planes $y=\pm D/2$ and an external magnetic field ${\bf H}$ is applied along the $z$-axis.

The current-phase relation of the Josephson junction is defined by the quasiparticle excitation energies $\varepsilon$ (hereinafter we consider the system of units with $\hbar=1$) \cite{Beenakker_Matrix}:
\begin{equation}\label{Current_Def}
I\left(\varphi\right)=-2e\sum\limits_{\varepsilon\in(0;\infty)}\frac{\partial \varepsilon}{\partial \varphi}\tanh\left(\frac{\varepsilon}{2T}\right).
\end{equation}
The energy levels $\varepsilon$ should be found from the solution of the BdG equations
\begin{equation}\label{BdG_Full}
\left(
  \begin{array}{cc}
    \hat{H} & \hat{\Delta} \\
    \hat{\Delta}^{\dag} & -\hat{H}^{\dag} \\
  \end{array}
\right)
\left(
    \begin{array}{c}
      u \\
      v \\
    \end{array}
\right)=
\varepsilon
\left(
    \begin{array}{c}
      u \\
      v \\
    \end{array}
\right).
\end{equation}
The electron- and hole-like parts of the quasiparticle wave function $u$ and $v$ are multicomponent:  $u=\left(u_{1\uparrow},u_{2\uparrow},u_{1\downarrow},u_{2\downarrow}\right)$ and $v=\left(v_{1\uparrow},v_{2\uparrow},v_{1\downarrow},v_{2\downarrow}\right)$, where the first indices enumerate the conductive channels and arrows indicate the $z$-axis spin projections. In Eq.~(\ref{BdG_Full}) $\hat{\Delta}$ is the superconducting pairing potential and $\hat{H}$ is the single-electron $4\times 4$-matrix Hamiltonian of the isolated wire, which for zero magnetic field takes the form
\begin{equation}\label{Ham_No_Field}
\hat{H}=\left[\xi\left(\hat{p}\right)-\mu+\alpha\hat{p} \hat{\sigma}_z\right]\otimes\hat{I}+\hat{V}(x).
\end{equation}
Here $\hat{p}=-i\partial_x$ is the $x$-projection of the momentum, $\xi\left(p\right)$ is the electron energy in the isolated wire, $\mu$ is the chemical potential, the term $\alpha\hat{p} \hat{\sigma}_z$ describes the Rashba spin-orbit interaction  originating from the broken inversion symmetry in the $y$-direction \cite{Rashba}, the operator $\hat{I}$ is a $2\times 2$ unit matrix in the channel subspace, and the potential $\hat{V}(x)$ describes the scattering at the S/nanowire interfaces. Applying the magnetic field we should include the Zeeman term $g\mu_B H\hat{\sigma}_z$  into (\ref{Ham_No_Field})
and replace the momentum $\hat{p}$ with $\left(\hat{p}+|e|A_x/c\right)$
taking, e.g., the gauge $A_x(y)=-Hy$.

Our strategy is to find the quasiclassical solutions of Eq.~(\ref{BdG_Full}) inside the nanowire where both $\hat{\Delta}$ and $\hat{V}$ are zero and then match the solutions at the ends of the wire using phenomenological scattering matrices. The solvability conditions for this matching will give us the energy levels $\varepsilon$.
As a first step we derive the quasiclassical version of Eq.~(\ref{BdG_Full}) inside the wire. Taking, e.g., the functions $u_{1\uparrow}$ and $u_{2\uparrow}$ one can separate the fast oscillating exponential factor: $u_{n\uparrow}= \tilde{u}_{n\uparrow}^\pm e^{\pm ip_F^\pm x}$, where the Fermi momenta $p_F^+$ and $p_F^-$ for $p>0$ and $p<0$ are different in the presence of the spin-orbit coupling. Then from the BdG equation (\ref{BdG_Full}) with $\hat{\Delta}=0$, $\hat{V}=0$ and $H=0$ we find:
\begin{equation}\label{BdG_reduced}
\left[\xi\left(p_F^\pm\right)-\mu\pm\alpha p_F^\pm\right]\tilde{u}_{n\uparrow}^\pm \mp i\left[\xi'\left(p_F^\pm\right) \pm\alpha\right] \partial_x \tilde{u}_{n\uparrow}^\pm=\varepsilon \tilde{u}_{n\uparrow}^\pm,
\end{equation}
where $\xi'\left(p\right)\equiv \partial\xi/\partial p$. The Fermi momenta are defined by the equations $\xi\left(p_F^\pm\right)=\mu\mp\alpha p_F^\pm$. Assuming $\alpha$ to be small we find: $p_F^\pm\approx\left[1\mp \alpha/\xi'\left(p_F^0\right)\right]p_F^0$, where $\xi\left(p_F^0\right)=\mu$. Performing the straightforward derivation of equations for $u_{n\downarrow}^\pm$, $v_{n\uparrow}^\pm$ and $v_{n\downarrow}^\pm$ we obtain:
\begin{equation}\label{BdG_Quasiclassical}
\begin{array}{c}{\ds \mp i v_F^\pm\partial_x \tilde{u}_{n\uparrow}^\pm=\varepsilon \tilde{u}_{n\uparrow}^\pm, ~~~~~\mp i v_F^\pm\partial_x \tilde{v}_{n\downarrow}^\pm=-\varepsilon \tilde{v}_{n\downarrow}^\pm}\\{\ds \mp i v_F^\mp\partial_x \tilde{u}_{n\downarrow}^\pm=\varepsilon \tilde{u}_{n\downarrow}^\pm, ~~~~~\mp i v_F^\mp\partial_x \tilde{v}_{n\uparrow}^\pm=-\varepsilon \tilde{v}_{n\uparrow}^\pm.}\end{array}
\end{equation}
Using the expansion $\xi'\left(p_F^\pm\right)= \xi'\left(p_F^0\right)\mp\alpha p_F^0\xi''\left(p_F^0\right)/\xi'\left(p_F^0\right)$, we find the Fermi velocities:
\begin{equation}\label{VF_renorm}
v_F^\pm=\xi'\left(p_F^0\right)\pm\alpha\left[1-p_F^0\xi''\left(p_F^0\right)/\xi'\left(p_F^0\right)\right].
\end{equation}
Clearly the spin-orbit coupling results in the difference between the Fermi velocities $v_F^+$ and $v_F^-$ of quasiparticles with opposite momenta. This renormalization (\ref{VF_renorm}) is absent only for exactly quadratic spectrum.
It is the difference between $v_F^+$ and $v_F^-$ which is responsible for the so-called $\varphi_0$-junction formation (see \cite{Buzdin_Phi} and discussion below). Thus, the above derivation expains the results of \cite{NoPhi0}, where no $\varphi_0$-junction was found for $\xi(p)\propto p^2$ spectrum, and the subsequent misinterpretation for the conditions of the $\varphi_0$-junction emergence in \cite{WrongPhi0}.

It is convenient to introduce the 4-component envelope wave functions $w^\pm_\sigma(x)=(\sqrt{v_F^\pm}\tilde{u}_{1\sigma}^\pm ,\sqrt{v_F^\pm}u_{2\sigma}^{\pm},\sqrt{v_F^\mp}v_{1-\sigma}^{\mp},\sqrt{v_F^\mp}v_{2-\sigma}^{\mp})$.
Considering, e.g., $w^\pm_\uparrow$ and neglecting the spin flip at the wire ends one can write the matching conditions:
$w^\pm_\uparrow\left(\pm L/2\right)=\hat{T}^\pm w^\pm_\uparrow\left(\mp L/2\right)$, and $w^\mp_\uparrow\left(\pm L/2\right)=\hat{Q}^\pm w^\pm_\uparrow\left(\pm L/2\right)$,
where the unitary matrices $\hat{T}^\pm$ and $\hat{Q}^\pm$ describe the quasiparticle transmission along the wire channels and both normal and Andreev scattering processes at the wire ends.
The solvability condition for the above matching equations
\begin{equation}\label{MainEq}
\det\left[\hat{Q}^-\hat{T}^-\hat{Q}^+\hat{T}^+-\hat{1}\right]=0
\end{equation}
gives us  the eigenvalues of the BdG equations (\ref{BdG_Full}).
The eigenvalues for the opposite spin component can be obtained replacing $\alpha$ and $g$ by $-\alpha$ and $-g$.

The form of the matrices $\hat{T}^\pm$ is defined by the solution of Eq.~(\ref{BdG_Quasiclassical}) generalized for a non-zero magnetic field. Assuming  different g-factors $g_1$ and $g_2$ in different channels and introducing a dimensionless magnetic flux $\phi=HLD/\Phi_0$ we obtain:
\begin{equation}\label{T_Def}
\hat{T}^\pm=\left(
                  \begin{array}{cc}
                    e^{i\left(p_F^\pm+\varepsilon/v_F^\pm\right)L}\hat{M}^\pm & \hat{0} \\
                    \hat{0} & e^{-i\left(p_F^\mp-\varepsilon/v_F^\mp\right)L}\hat{M}^\mp \\
                  \end{array}
                \right).
\end{equation}
Here the $2\times 2$ matrices $\hat{M}^\pm$ have the elements $\hat{M}^\pm_{nl}=\exp\left[-ig_n\mu_B HL/v_F^\pm\mp(-1)^n i\pi\phi/2\right]\delta_{nl}$, where $\delta_{nl}$ is the Kronecker-delta. The phenomenological scattering matrices $\hat{Q}^\pm$ have a general form
\begin{equation}\label{Q_Def}
\hat{Q}^\pm=\left(
                  \begin{array}{cc}
                    \hat{R}_e^\pm & \hat{A}_h^\mp \\
                    \hat{A}_e^\pm & \hat{R}_h^\mp \\
                  \end{array}
                \right),
\end{equation}
where diagonal and off-diagonal elements are the $2\times 2$ matrices describing the normal and Andreev reflection from the S leads, respectively \cite{Unitarity}.

First, we consider the limit when the quasiparticles experience {\it full} Andreev reflection in each channel separately. The Andreev reflection is caused by the superconducting gap $\Delta_n$ induced in the $n$-th channel due to the proximity effect to the S leads. Note that for small tunneling rates $\Gamma_n$ between the $n$-th channel and the S lead the induced gap can be estimated as $\Delta_n\propto \Gamma_n$ \cite{Melnikov_Kopnin}. The above assumption of full Andreev reflection means that the size $d_s$ of the induced gap regions  well exceeds the relevant coherence length. In this limiting case the normal scattering vanish ($\hat{R}_e^\pm=\hat{R}_h^\pm=\hat{0}$) while the Andreev scattering is described by the matrices $(\hat{A}_e^\pm)_{nl}=\delta_{nl} \exp\left[\mp i\varphi/2-i\arccos(\varepsilon/\Delta_n)\right]$.

In the short junction limit ($\varepsilon L/v_F^\pm\ll 1$) only the sub-gap Andreev states contribute to the Josephson current.
Then solving Eq.~(\ref{MainEq}) and taking into account all spin projections we obtain four positive subgap energy levels
\begin{equation}\label{FullAndreev_Energy}
\varepsilon=\Delta_n\left|\cos\left[\varphi/2-(-1)^{n}\pi\phi/2\pm g_n\mu_BHL/v_F^\pm\right]\right|,
\end{equation}
where $n$ enumerates the channels.
For large temperatures $T\gg \Delta_n$ the current-phase relation (\ref{Current_Def}) takes the form
\begin{equation}\label{Current_GeneralAnswer}
I=\sum\limits_{n=1,2} I_n\sin\left[\varphi+\beta_nH+(-1)^n\pi\phi\right] \cos\left(\gamma_nH\right).
\end{equation}
Here $I_n=\left|e\right|\Delta_n^2/4T$ is the critical current of the $n$-th channel at $H=0$, the flux $\phi$ produces the  SQUID-like oscillations of $I_c$, the cosine term depending on the constants $\gamma_n=g_n\mu_BL\left(1/v_F^++1/v_F^-\right)$ describes the oscillatory behavior of $I_c$ due to the Zeeman interaction similar to the one in superconductor/ferromagnet/superconductor structures \cite{Buzdin_RMP}. The term $\beta_nH=g_n\mu_BLH\left(1/v_F^+-1/v_F^-\right)$ describes the $\varphi_0$-junction formation due to the spin-orbit coupling \cite{Buzdin_Phi}. The critical current corresponding to (\ref{Current_GeneralAnswer}) reads
\begin{equation}\label{CritCurr_Tc}
\begin{array}{c}{
I_c^2=I_1^2\cos^2\left(\gamma_1H\right)+I_2^2\cos^2\left(\gamma_2H\right)}\\{+2I_1I_2\cos\left(\gamma_1H\right)\cos\left(\gamma_2H\right) \cos\left[2\pi\phi+\left(\beta_1-\beta_2\right)H\right].}\end{array}
\end{equation}
Interestingly in case of the difference between $g$-factors in the conducting channels the spin-orbit coupling influences the period of the SQUID-like orbital oscillations in $I_c(H)$, i.e. renormalizes the effective quantization area enclosed by the channels: $S_{eff}=LD+\Phi_0(\beta_1-\beta_2)/2\pi$.
%%%%%%%%%%%%%%%%%%%%%%%%%%%%%%%%%%%%%%%%%%%%%
\begin{figure}[b!]
\includegraphics[width=0.48\textwidth]{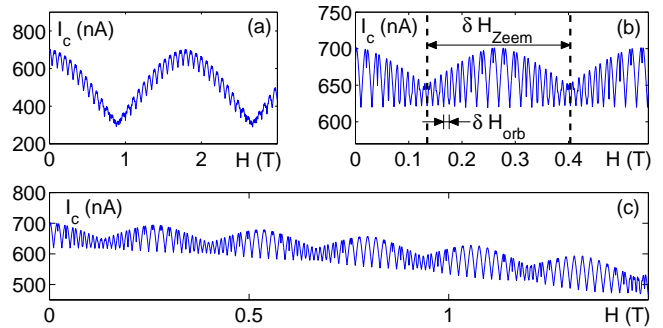}
\caption{(Color online) The critical current $I_c$ vs the magnetic field $H$. We choose $T=0.1~{\rm K}$, $\Delta_1=7.5~ {\rm K}$, $\Delta_2=1~ {\rm K}$, $v_F=3\cdot 10^5~ {\rm m/s}$, $L=2~{\rm \mu m}$ and (a) $D=15~{\rm nm}$ and (b)-(c) $D=50~{\rm nm}$. We also take (a) $g_1=g_2=1.5$; (b) $g1=0$ and $g_2=10$; (c) $g1=1$ and $g_2=10$.} \label{Fig_IcH}
\end{figure}
%%%%%%%%%%%%%%%%%%%%%%%%%%%%%%%%%%%%%
%
%
Choosing the parameters relevant to the experimental situation in \cite{Bouchiat_ArXiv} we obtain a variety of $I_c(H)$ dependencies shown in Fig.~\ref{Fig_IcH}. These dependencies reproduce not only multi-periodic oscillations due to the interplay of the orbital and Zeeman interactions observed in \cite{Bouchiat_ArXiv} but also asymmetry in the form of the upper and lower envelopes. In Fig.~\ref{Fig_IcH}(a)-(b) one can clearly see two periods of oscillations: $\delta H_{orb}=\Phi_0/S_{eff}$ and $\delta H_{Zeem}=2\pi/\gamma_1=2\pi/\gamma_2$. The slow drift of the average current in Fig.~\ref{Fig_IcH}(d) should be considered in fact as a fragment of the large-period oscillations caused by the difference between $\gamma_1$ and $\gamma_2$.

Now let us study the crossover between the limits of large and small Andreev reflection which occurs with the decrease in the induced gap value. We neglect for simplicity the spin-orbit coupling, the Zeeman interaction and the difference between the induced gaps ($\Delta_1=\Delta_2\equiv \Delta_0$). We assume the inter-channel electron transfer to be the only normal scattering mechanism at the ends of the nanowire (in the opposite limit of vanishing inter-channel transfer the current-phase relation should be similar to the one for a quantum box studied in \cite{Kopnin_Melnikov}).
Thus, we take the scattering matrices in the form $(\hat{R}_{e,h}^\pm)_{nl}=t(1-\delta_{nl})$ and $(\hat{A}_{e,h}^\pm)_{nl}=a\delta_{nl}e^{\mp i\varphi/2}$, where $a=-i\Delta_0\sinh(qd_s)/Z$,  $q=\sqrt{\Delta_0^2-\varepsilon^2}/v_F$, $t=qv_F/Z$, and $Z=qv_F\cosh(qd_s)+i\varepsilon\sinh(qd_s)$ \cite{Demers}. The inter-channel hopping with the amplitude $t$ allows the formation of closed electron orbits of non-zero area and, thus, can strongly affect the electron transfer through the nanowire due to the interference between the channels. Such model provides the simplest way to clarify if these closed orbits can cause the interplay between $2\Phi_0$ and $\Phi_0$ flux periodicities in the critical current corresponding to the AB interference of electrons and Cooper pairs.

\begin{figure}[b!]
\includegraphics[width=0.48\textwidth]{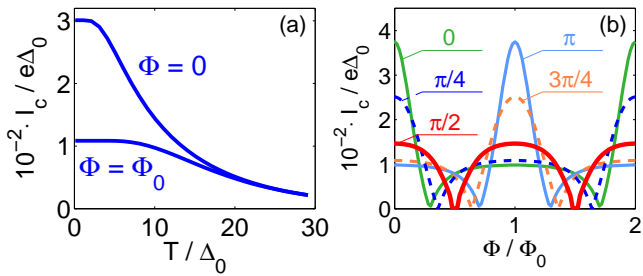}
\caption{(Color online) (a) The temperature crossover from $2\Phi_0$- to $\Phi_0$-periodic oscillations of the critical current $I_c$. The curves correspond to $\Phi=0$ and $\Phi=\Phi_0$. We take $L=d_s=0.01v_F/\Delta_0$ and $p_FL_0=\pi/4+2\pi m$ ($m$ is an integer number). (b) Dependencies $I_c(\Phi)$ for $T/\Delta_0=5$ and $p_FL_0=\eta+2\pi m$ where the values $\eta$ are shown near the curves.} \label{Fig_3}
\end{figure}

In Fig.~\ref{Fig_3} we present the results of the critical current calculations for the energy spectrum given by Eq.~(\ref{MainEq}) \cite{supp}. Generally the period of $I_c(H)$ oscillations strongly depends both on temperature $T$ and the parameter $p_FL_0=2p_F(L+2d_s)$ controlling mesoscopic fluctuations. In the limit $d_s\gg v_F/\Delta_0$ we get the case of independent channels considered above and restore the $\Phi_0$-periodicity of the $I_c(H)$ oscillations.
Substantial difference between the curves $I_c(T)$ for $\Phi=0$ and $\Phi=\Phi_0$ appears only for $d_s<v_F/\Delta_0$. In this regime the Andreev reflection is week and one can clearly see the $\Phi_0$ - $2\Phi_0$ crossover. For low temperatures $T<v_F/L_0$ the curves in Fig.~\ref{Fig_3} are strongly different since the system transparency and the corresponding critical current oscillate with the electron AB period $2\Phi_0$. For higher temperatures the normal metal coherence length $v_F/T$ can become less than the length $L_0$ of the closed electron path and the $2\Phi_0$-periodic interference of electrons can not contribute to the superflow through the junction. Thus, with the temperature increase ($T>v_F/L_0$) the difference between curves in Fig.~\ref{Fig_3} vanishes and $I_c$ oscillates with the AB period of Cooper pairs ($\Phi_0)$.

%%%%%%%%%%%%%%%%%%%%%%%%%%%%%%%%%%%%%%%%%%%%%
%%%%%%%%%%%%%%%%%%%%%%%%%%%%%%%%%%%%%

At temperatures close to $T_c$ it is quite natural to expect that the system behavior can be well understood within the  Ginzburg-Landau  approach modified to include the Zeeman and spin-orbit interactions. Keeping only the terms of the order $O(\psi^2)$ we consider the free energy density $F$ in the form \cite{Samokhin,Kaur}
\begin{equation}\label{GL_F}
\begin{array}{c}{\ds
F=\sum_{n=1,2}\left\{ a\left|\Psi_n\right|^2+\gamma\left|\hat{D}_x\Psi_n\right|^2+\beta\left|\hat{D}_x^2\Psi_n\right|^2\right.}\\{\ds \left.-\nu H \left[\Psi_n\left(\hat{D}_x\Psi_n\right)^*+\Psi_n^*\left(\hat{D}_x\Psi_n\right)\right]\right\},}\end{array}
\end{equation}
where $\Psi_n$ is the superconducting order parameter in the $n$-th channnel, $a(x)\sim\left[T-T_c(x)\right]$ and inside the nanowire $a>0$, $\hat{D}_x=-i\partial_x+2\pi A_x/\Phi_0$ and the constant $\nu\sim\alpha g$ describes the strength of the spin-orbit coupling. The oscillatory behavior of the Cooper pair wave function due to the Zeeman interaction reveals only for the magnetic fields above the tricritical Lifshitz point, i.e., for $\gamma<0$ \cite{Buzdin_RMP}. Accounting for the higher order gradient term with $\beta>0$ in (\ref{GL_F}) one can find an additional characteristic length scale $\xi_f=\sqrt{\beta/\left|\gamma\right|}$ corresponding to the period of the gap function oscillation in the Fulde-Ferrell-Larkin-Ovchinnikov phase. The Josephson current for  $\gamma=0$ has been previously calculated in \cite{Buzdin_Kulic}. Here we   analyze the case of arbitrary negative $\gamma$ values restricting ourselves by the condition $\xi_f<\xi=\sqrt{|\gamma|/a}$ meaning the absence of the intrinsic superconductivity in the nanowire. For simplicity we also make several assumptions: (i) the spin-orbit coupling is weak and can be treated perturbatively; (ii) $L\gg\sqrt{\xi^2+\xi_f\xi}$; (iii) inside the S leads the Zeeman interaction is negligible; (iv) the conductivity of the S leads strongly exceeds the one in the nanowire so the inverse proximity effect in the S leads can be neglected; (v) at the S/nanowire interfaces there is no barrier and, as a consequence, the order parameter is continuous at $x=\pm L/2$: $\Psi_n\left(\pm L/2\right)=\Delta_n \exp\left(\pm i\varphi/2\right)$.
Using the boundary conditions we find the supercurrent $j_x=-c\delta F/\delta A_x$
in the $n$-th channel  \cite{supp}:
$j_n=j_c^{(n)} \sin\left[\varphi+(-1)^n\pi \Phi/\Phi_0+\varphi_0\right]$,
where $\sin\varphi_0=\sinh(sL)\cos\chi/\sqrt{\sin^2\chi+\sinh^2(sL)}$,
\begin{equation}\label{GL_Curr_Res}
j_c^{(n)}=\frac{16\left|e\right|\beta\Delta_n^2 k^-}{(\xi_f\xi)^{3/2}k^+}e^{-\frac{k^-L}{\sqrt{2}}} \sqrt{\sin^2\chi+\sinh^2(sL)} \ ,
\end{equation}
${\sin (\chi-k^+L/\sqrt{2})=\left(1-2\xi/\xi_f\right)k^+\sqrt{\xi_f\xi/2}}$,
 ${s=\nu H/(2\beta k^+k^-)}$, $k^\pm=\xi^{-1}\sqrt{\xi/\xi_f\pm1}$.

Summing up the contributions from both channels we find the magnetic field dependence of the critical current demonstrating the multi-periodic magnetic oscillations. The period of the fast oscillations is again equal to $\Phi_0/LD$ while the slow oscillations caused by the Zeeman interaction are determined by the dependence of the coefficient $\gamma$ on $H$. For long junctions with $L\sim s^{-1}$ the term $\sinh^2(sL)$ can result in the increase in $I_c$ with the increasing $H$. Obviously this effect can be suppressed because of damping of the superconductivity inside the S leads due to the magnetic field. However for the Pb films and ${\rm LaAlO_3/SrTiO_3}$ heterostructures with strong spin-orbit coupling in rather small magnetic fields the increasing dependencies $T_c(H)$ were observed \cite{IncreasingTc}. In this case as follows from (\ref{GL_Curr_Res}) the dependencies $I_c(H)$ should reveal the increasing trend due to the spin-orbit coupling.

To sum up we have investigated the distinctive features of the very rich interference physics in nanowires coupled to the superconducting leads and suggest phenomenological models explaining the multi-periodic magnetic oscillations in supercurrent through these systems.

The authors thank H. Bouchiat and A. Murani for stimulating discussions. This work was
supported by the French
ANR ``MASH," NanoSC COST Action MP1201, the Russian Foundation for Basic Research,
the Russian Presidential foundation (Grant SP-6340.2013.5), and the grant of the Russian Ministry of Science and Education No.02.B.49.21.0003.

\end{document}